\documentstyle[prd,preprint,tighten,aps,amssymb,newlfont,epsfig]{revtex}
\begin{document}
\draft
\preprint{
\begin{tabular}{r}
DFTT 69/97\\
UWThPh-1997-45\\
IASSNS-AST 97/67\\
hep-ph/9711432
\end{tabular}
}
\title{Constraints on long-baseline neutrino oscillations from
the results of neutrino oscillation experiments}
\author{S.M. Bilenky}
\address{Joint Institute for Nuclear Research, Dubna, Russia, and\\
Institute for Advanced Study, Princeton, N.J. 08540}
\author{C. Giunti}
\address{INFN, Sezione di Torino, and Dipartimento di Fisica Teorica,
Universit\`a di Torino,\\
Via P. Giuria 1, I--10125 Torino, Italy}
\author{W. Grimus}
\address{Institute for Theoretical Physics, University of Vienna,\\
Boltzmanngasse 5, A--1090 Vienna, Austria}
\maketitle
\begin{abstract}
It is shown that
in the two schemes with four massive neutrinos
which are compatible with
the results of all neutrino oscillation experiments,
the probabilities of
$\bar\nu_e$
disappearance and
$
\stackrel{\makebox[0pt][l]
{$\hskip-3pt\scriptscriptstyle(-)$}}{\nu_{\mu}}
\to\stackrel{\makebox[0pt][l]
{$\hskip-3pt\scriptscriptstyle(-)$}}{\nu_{e}}
$
appearance in long-baseline experiments
are severely constrained.
\end{abstract}

\pacs{Talk presented by C. Giunti at the
Fifth International Workshop on
Topics in Astroparticle and Underground Physics,
TAUP97,
September 7--11, 1997,
Laboratori Nazionali del Gran Sasso, Assergi, Italy.}

The problem of the existence of neutrino masses and mixing is
considered today one of the most important in high-energy physics
and many experiments are dedicated to it.
Among the numerous existing experimental results
there are three positive indications which come from
neutrino oscillation experiments.
Neutrino oscillations
can occur only if neutrinos are massive particles,
if their masses are different
and if neutrino mixing is realized in nature
(see \cite{BP78}).
In this case,
the left-handed flavor neutrino
fields
$\nu_{{\alpha}L}$
are superpositions
of
the left-handed components
$\nu_{kL}$
of the fields of neutrinos with definite mass
($k=1,2,3,\ldots$):
$
\nu_{{\alpha}L}
=
\sum_k
U_{{\alpha}k}
\,
\nu_{kL}
$,
where $U$
is a unitary mixing matrix.

The three experimental indications
in favor of neutrino oscillations come from the anomalies observed in
the solar neutrino experiments \cite{sunexp},
in the atmospheric neutrino experiments \cite{atmexp}
and in the LSND experiment \cite{LSND}.
The solar neutrino deficit can be explained
by transitions of $\nu_e$'s
into other states
due to a mass-squared difference
of the order of
$ 10^{-5} \, \mathrm{eV}^2 $
in the case of resonant MSW transitions
or
$ 10^{-10} \, \mathrm{eV}^2 $
in the case of vacuum oscillations.
The atmospheric neutrino anomaly
can be explained
by transitions of $\nu_\mu$'s
into other states
due to a mass-squared difference
of the order of
$ 10^{-2} \, \mathrm{eV}^2 $.
Finally,
the LSND experiment found
indications in favor of
$ \bar\nu_\mu \to \bar\nu_e $
oscillations
due to a mass-squared difference
from
$ 0.3 \, \mathrm{eV}^2 $
to
$ 2.2 \, \mathrm{eV}^2 $
(this range takes into account
the negative results of other short-baseline experiments).

Hence,
three different scales of mass-squared difference
are needed in order to explain the three indications
in favor of neutrino oscillations.
This means that the number of massive neutrinos
must be bigger than three.
In the following we consider the simplest possibility
of existence of four massive neutrinos ($n=4$).
In this case,
besides the three light flavor neutrinos
$\nu_e$,
$\nu_\mu$,
$\nu_\tau$
that
contribute to the invisible width of the $Z$-boson
measured with high accuracy by LEP experiments,
there is a light sterile flavor neutrino
$\nu_s$
that does not take part in
the standard weak interactions.

In \cite{BGG96} we have shown that among all the possible
four-neutrino mass spectra
only two are compatible
with the results of all neutrino oscillation experiments:
\begin{eqnarray}
\mathrm{A}.
&\qquad&
\underbrace{
\overbrace{m_1 < m_2}^{\mathrm{atm}}
\ll
\overbrace{m_3 < m_4}^{\mathrm{sun}}
}_{\mathrm{LSND}}
\,,
\label{01}
\\
\mathrm{B}.
&\qquad&
\underbrace{
\overbrace{m_1 < m_2}^{\mathrm{sun}}
\ll
\overbrace{m_3 < m_4}^{\mathrm{atm}}
}_{\mathrm{LSND}}
\,.
\label{02}
\end{eqnarray}
In these two schemes
the four neutrino masses
are divided in two pairs of close masses
separated by a gap of about 1 eV,
which provides the mass-squared difference
$ \Delta{m}^2 \equiv \Delta{m}^2_{41} \equiv m_4^2 - m_1^2 $
that is relevant for the oscillations
observed in the LSND experiment.
In scheme A,
$ \Delta{m}^{2}_{21} \equiv m_2^2 - m_1^2 $
is relevant
for the explanation of the atmospheric neutrino anomaly
and
$ \Delta{m}^{2}_{43} \equiv m_4^2 - m_3^2 $
is relevant
for the suppression of solar $\nu_e$'s.
In scheme B,
the roles of
$\Delta{m}^{2}_{21}$
and
$\Delta{m}^{2}_{43}$
are reversed.

The results of all neutrino oscillation experiments
are compatible with the schemes A and B
if
(see \cite{BGG96})
\begin{equation}
c_e \leq a_e^0
\quad \mbox{and} \quad
c_\mu \geq 1 - a_\mu^0
\,,
\label{03}
\end{equation}
where the quantities $c_\alpha$,
with $\alpha=e,\mu$,
are defined as
\begin{equation}
c_\alpha
\equiv
\sum_{k}
|U_{{\alpha}k}|^2
\,,
\label{04}
\end{equation}
with
$k=1,2$ in scheme A
and
$k=3,4$ in scheme B.
The quantities 
$a_e^0$ and $a_\mu^0$
depend on $\Delta{m}^2$ and
are derived from the exclusion plots of
short-baseline (SBL)
reactor and accelerator disappearance experiments.
The exclusion curves obtained in the Bugey reactor experiment
and in the CDHS and CCFR accelerator experiments
\cite{Bugey95-CDHS84-CCFR84}
imply that both
$a_e^0$ and $a_\mu^0$
are small quantities:
$ a^{0}_e \lesssim 4 \times 10^{-2} $
and
$ a^{0}_\mu \lesssim 2 \times 10^{-1} $
for any value of
$\Delta{m}^{2}$
in the range
$
0.3
\lesssim
\Delta{m}^2
\lesssim
10^{3} \, \mathrm{eV}^2
$
(see Fig.1 of Ref.\cite{BBGK96}).

The smallness of $c_e$
in both schemes A and B,
which is a consequence of the solar neutrino problem,
implies that the electron neutrino has a
small mixing with the neutrinos whose mass-squared difference is
responsible for the oscillations of atmospheric neutrinos
(i.e.,
$\nu_1$, $\nu_2$ in scheme A and $\nu_3$, $\nu_4$ in scheme
B).
Hence,
the transition probability of
electron neutrinos and antineutrinos
into other states
in atmospheric and long-baseline (LBL) experiments
is suppressed.
Indeed,
as shown in \cite{BGG97},
the transition probabilities
of electron neutrinos and antineutrinos
into all other states are bounded by
\begin{equation}
1 -
P^{(\mathrm{LBL})}_{\stackrel{\makebox[0pt][l]
{$\hskip-3pt\scriptscriptstyle(-)$}}{\nu_{e}}
\to\stackrel{\makebox[0pt][l]
{$\hskip-3pt\scriptscriptstyle(-)$}}{\nu_{e}}}
\leq
a^{0}_{e}
\left( 2 - a^{0}_{e} \right)
\,.
\label{05}
\end{equation}
The curve corresponding
to this limit
obtained from the 90\% CL exclusion plot of the Bugey
experiment is shown
in Fig.\ref{fig1}
(solid line). For comparison,
the expected sensitivity
of the LBL reactor neutrino experiments
CHOOZ and Palo Verde \cite{CHOOZ-PV}
are also shown in
Fig.\ref{fig1}
by the dashed and dash-dotted vertical lines, respectively.
The shadowed region in Fig.\ref{fig1}
corresponds to the range of $\Delta{m}^2$
allowed at 90\% CL by the results of the LSND
and all the other SBL experiments.
It can be seen that
the LSND signal indicates an upper bound for
$1-P^{(\mathrm{LBL})}_{\bar \nu_e \to \bar \nu_e}$
of about
$ 5 \times 10^{-2} $,
smaller than the expected sensitivities of
the CHOOZ and Palo Verde experiments.

For the probability of
$
\stackrel{\makebox[0pt][l]
{$\hskip-3pt\scriptscriptstyle(-)$}}{\nu_{\mu}}
\to\stackrel{\makebox[0pt][l]
{$\hskip-3pt\scriptscriptstyle(-)$}}{\nu_{e}}
$
transitions in LBL experiments we have \cite{BGG97}
\begin{equation}
P^{(\mathrm{LBL})}_{\stackrel{\makebox[0pt][l]
{$\hskip-3pt\scriptscriptstyle(-)$}}{\nu_{\mu}}
\to\stackrel{\makebox[0pt][l]
{$\hskip-3pt\scriptscriptstyle(-)$}}{\nu_{e}}}
\leq
\min\!\left(
a^{0}_{e}
\left( 2 - a^{0}_{e} \right)
\, , \,
a^{0}_{e}
+
\frac{1}{4}
\,
A^{0}_{\mu;e}
\right)
,
\label{06}
\end{equation}
where
$A^{0}_{\mu;e}$
is the upper bound for the amplitude of
$
\stackrel{\makebox[0pt][l]
{$\hskip-3pt\scriptscriptstyle(-)$}}{\nu_{\mu}}
\to\stackrel{\makebox[0pt][l]
{$\hskip-3pt\scriptscriptstyle(-)$}}{\nu_{e}}
$
transitions measured in SBL experiments with accelerator neutrinos.
The bound obtained with Eq.(\ref{06})
from the 90\% CL exclusion plots of the Bugey
experiment
and
of the
BNL E734, BNL E776 and CCFR
experiments \cite{BNLE734-BNLE776-CCFR96}
is depicted by the dashed line
in Fig.\ref{fig2}.
The solid line in Fig.\ref{fig2} shows the upper bound on
$
P^{(\mathrm{LBL})}_{\stackrel{\makebox[0pt][l]
{$\hskip-3pt\scriptscriptstyle(-)$}}{\nu_{\mu}}
\to\stackrel{\makebox[0pt][l]
{$\hskip-3pt\scriptscriptstyle(-)$}}{\nu_{e}}}
$
taking into account matter effects.
The expected sensitivities
of the K2K long-baseline accelerator neutrino experiment
\cite{K2K} is indicated in Fig.\ref{fig2}
by the dash-dotted vertical line.
The shadowed region in Fig.\ref{fig1}
corresponds to the range of $\Delta{m}^2$
allowed at 90\% CL by the results of the LSND
and all the other SBL experiments.
It can be seen that the results of SBL experiments
indicate an upper bound for
$
P^{(\mathrm{LBL})}_{\stackrel{\makebox[0pt][l]
{$\hskip-3pt\scriptscriptstyle(-)$}}{\nu_{\mu}}
\to\stackrel{\makebox[0pt][l]
{$\hskip-3pt\scriptscriptstyle(-)$}}{\nu_{e}}}
$
smaller than
$ 4 \times 10^{-2} $
and
smaller than the expected sensitivity of
the K2K experiment.

In conclusion,
we have shown that
in the four-neutrino schemes A and B
which are compatible with
the results of all neutrino oscillation experiments,
the probabilities of
$\bar\nu_e$
disappearance and
$
\stackrel{\makebox[0pt][l]
{$\hskip-3pt\scriptscriptstyle(-)$}}{\nu_{\mu}}
\to\stackrel{\makebox[0pt][l]
{$\hskip-3pt\scriptscriptstyle(-)$}}{\nu_{e}}
$
appearance in LBL experiments
are severely constrained\footnote{After
we finished this paper the results of the first
LBL reactor experiment CHOOZ
appeared \cite{CHOOZ97}.
No indications in favor of
$\bar\nu_e\to\bar\nu_e$
transitions were found in this experiment.
The upper bound for the
probability
$ 1 - P^{(\mathrm{LBL})}_{\bar\nu_e\to\bar\nu_e} $
found in the CHOOZ experiment is in agreement with
the limit presented in Fig.\ref{fig1}.}
(on the other hand,
the channels
$
\stackrel{\makebox[0pt][l]
{$\hskip-3pt\scriptscriptstyle(-)$}}{\nu_{\mu}}
\to\stackrel{\makebox[0pt][l]
{$\hskip-3pt\scriptscriptstyle(-)$}}{\nu_{\mu}}
$
and
$
\stackrel{\makebox[0pt][l]
{$\hskip-3pt\scriptscriptstyle(-)$}}{\nu_{\mu}}
\to\stackrel{\makebox[0pt][l]
{$\hskip-3pt\scriptscriptstyle(-)$}}{\nu_{\tau}}
$
are not constrained at all).

\newpage

\begin{minipage}{0.95\textwidth}
\begin{center}
\mbox{\epsfig{file=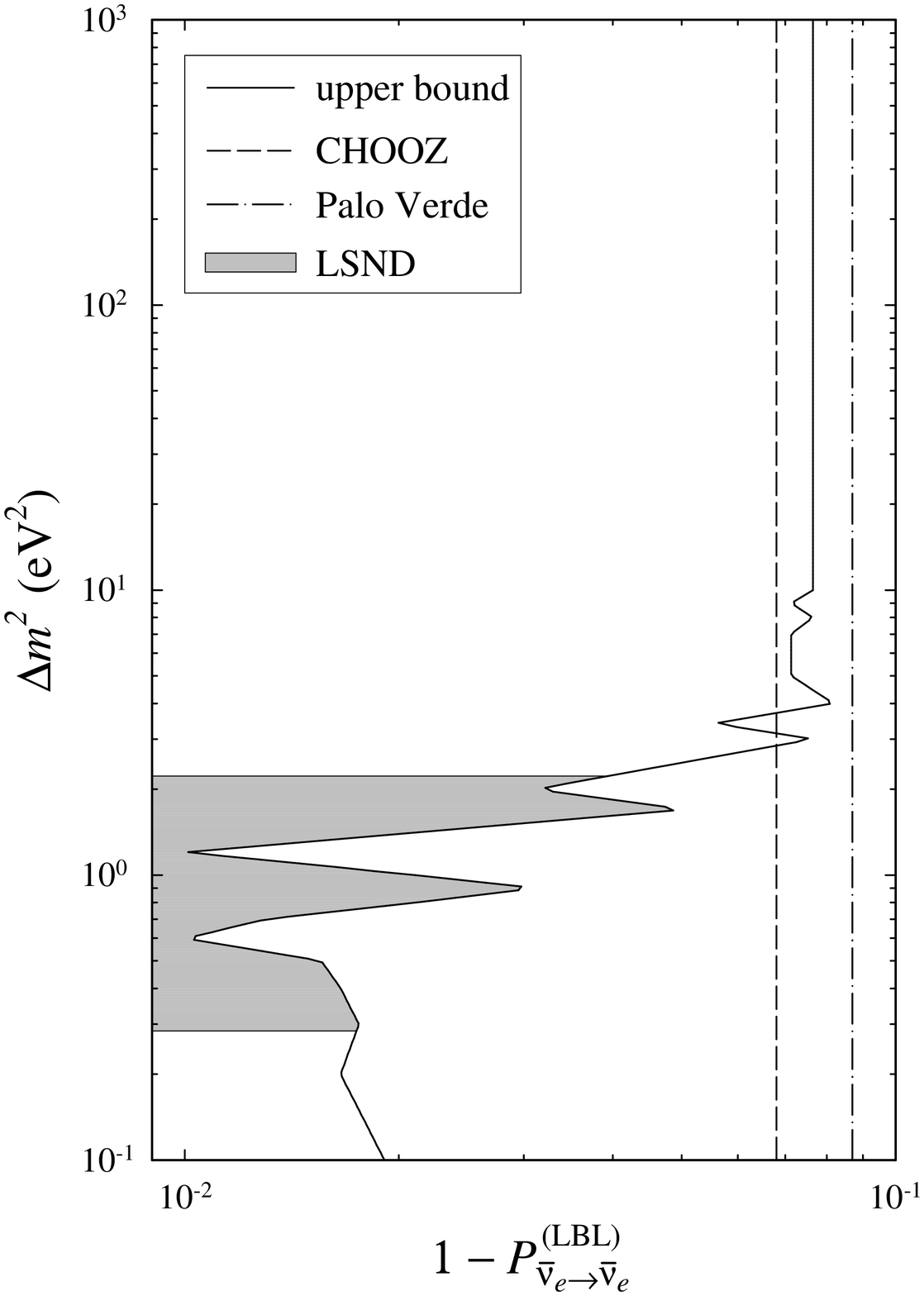,height=0.95\textheight}}
\end{center}
\end{minipage}
\begin{center}
\refstepcounter{figure}
\label{fig1}                 
\Large Figure \ref{fig1}
\end{center}

\newpage

\begin{minipage}{0.95\textwidth}
\begin{center}
\mbox{\epsfig{file=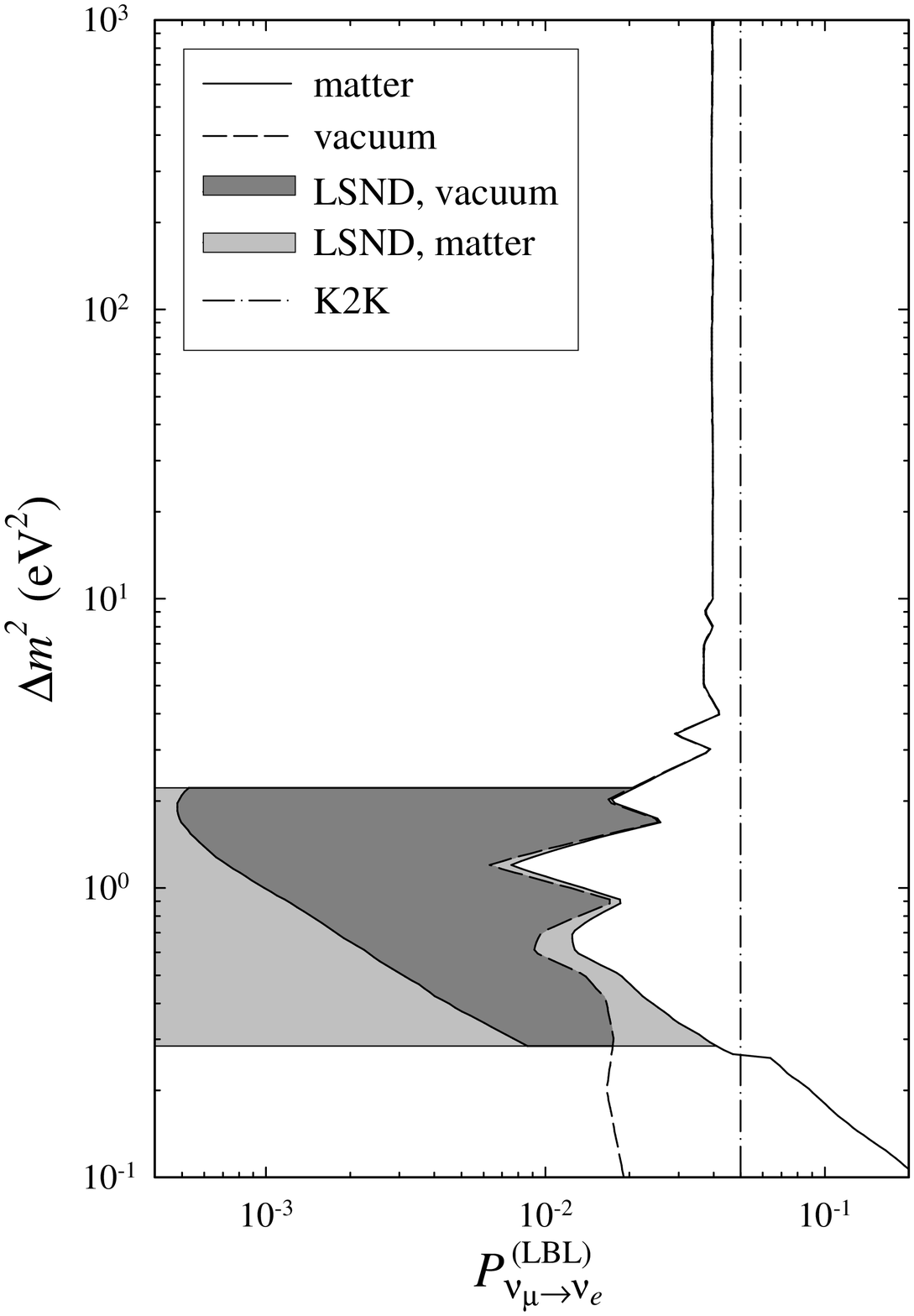,height=0.95\textheight}}
\end{center}
\end{minipage}
\begin{center}
\refstepcounter{figure}
\label{fig2}                 
\Large Figure \ref{fig2}
\end{center}

\end{document}